\documentclass{article}

\usepackage{makeidx}     % allows index generation
\usepackage{graphicx}    % standard LaTeX graphics tool
\usepackage{multicol}    % used for the two-column index
\usepackage{amsmath}
\usepackage{amsfonts}
\usepackage{amsthm}
\usepackage{subfigure}
\makeindex             % used for the subject index
\renewcommand{\d}{\mathrm{d}}
\newcommand{\e}{\mathrm{e}}

\newtheorem{theorem}{Theorem}

\title{Damping factors for the gap-tooth scheme}
\author{Giovanni Samaey, %\inst{1}\and 
Ioannis G. Kevrekidis,  % \inst{2} \and 
and Dirk Roose
}
\begin{document}

%\titlerunning{Damping factors for the gap-tooth scheme} %for an abbreviated version of

%\inst{1}}
%\institute{Department of Computer Science, K.U. Leuven, Celestijnenlaan 200A, 3000 Leuven,
% Belgium.
%\texttt{giovanni.samaey@cs.kuleuven.ac.be}
%\and Department of Chemical Engineering, PACM and Department of Mathematics,
%Princeton University, Princeton, USA.
%\texttt{yannis.kevrekidis@princeton.edu}
%\and Department of Computer Science, K.U. Leuven, Celestijnenlaan 200A, 3000 Leuven,
%Belgium. \texttt{dirk.roose@cs.kuleuven.ac.be}
\maketitle

\begin{abstract}An important class of problems exhibits macroscopically smooth behaviour
in space and time, while only a microscopic evolution law is known.  
For such time-dependent multi-scale problems, the gap-tooth scheme has recently been 
proposed.  
The scheme approximates the evolution of an unavailable (in closed form) macroscopic
equation in a macroscopic domain; it only uses appropriately initialized simulations 
of the available microscopic model in a number of small boxes. 
For some model problems, including numerical homogenization, the scheme is
essentially equivalent to a finite difference scheme, provided we repeatedly 
impose appropriate algebraic constraints on the solution for each box.
%
%We have proved %{\bf neeed a reference - or is it here ?} 
%that the scheme is essentially equivalent
%to a finite difference scheme for the macroscopic equation for some model
%problems, including numerical homogenization, when we appropriately
%constrain the solution near the boundaries of the boxes.
%
Here, we demonstrate that it is possible to obtain a convergent scheme without 
constraining the microscopic code, by introducing buffers that ``shield"
over relatively short times the dynamics
inside each box from boundary effects.  
We explore and quantify the behavior of these schemes 
systematically through the numerical computation of damping factors 
of the corresponding coarse time-stepper, for
which no closed formula is available.
\end{abstract}

\section{Introduction}
\label{sec:1}
For an important class of multi-scale problems, a separation
of scales exists between the (microscopic, detailed) level of description of
the available model, and the (macroscopic, continuum) level at which one
would like to observe the system.  
Consider, for example,
a kinetic Monte Carlo model of bacterial chemotaxis \cite{SetGearOthKevr03}. 
A stochastic biased random walk model describes the probability of
an individual bacterium to run or ``tumble", based on the rotation of its flagellae.
Technically, it would be possible to run the detailed model 
for all space and
time, and observe the macroscopic variables of interest, 
but this would be prohibitively
expensive.
It is known, however, that, under certain conditions, one can write a closed deterministic 
model for the evolution
of the \emph{concentration} of the bacteria as a function of space and
time.

The recently proposed \emph{gap-tooth scheme} 
\cite{KevrGearHymKevrRunTheo02} can then be used instead of performing 
stochastic time integration in the 
whole domain.
A number of small intervals (boxes, \emph{teeth}), separated by large gaps, are introduced;
they qualitatively correspond to mesh points for a traditional, continuum solution of the
unavailable chemotaxis equation.
The scheme works as  follows: We first choose a number of macroscopic grid points.  
We then choose a small interval around
each grid point; initialize the fine scale, microsopic solver
within each interval consistently with the macroscopic initial 
conditions;
and provide each box with appropriate (as we will see, to some extent artificial)
boundary conditions.  
Subsequently, we use the microscopic model
in each interval to simulate evolution until time $\Delta t$,
and obtain macroscopic information (e.g. by computing the average 
density in each box) at time $\Delta t$.  
This amounts to a coarse time-$\Delta t$ map; this procedure is then repeated.

The generalized Godunov scheme of E and Engquist \cite{EEng03}
also solves an unavailable macroscopic equation by repeated calls to a microscopic code;
however, the assumption is made that the unavailable equation can be written in conservation form.
In the gap-tooth scheme discussed here, the microscopic computations are
performed without assuming such a form for the ``right-hand-side" of the
unavailable macroscopic equation;  we evolve the detailed
model in a subset of the domain, and try to recover macroscopic evolution
through interpolation in space and extrapolation in time.

We have showed analytically, in the context of numerical homogenization, that the
gap-tooth scheme is close to a finite difference scheme for the 
homogenized equation \cite{SamKevrRoo03}.  
However, that analysis employed simulations using an algebraic constraint, 
ensuring that the initial macroscopic gradient
is preserved at the boundary of each box over the time-step $\Delta t$.  
This requires altering an existing microscopic code, so as to impose this 
macroscopically-inspired constraint.  
This may be impractical (e.g.\ if the macroscopic gradient has to be estimated), 
undesirable 
(e.g.\ if the development of the code is expensive and time-consuming) or even
impossible (e.g.\ if the microscopic code is a \emph{legacy code}).
Generally, a given microscopic code 
allows us to run with a set of pre-defined boundary
conditions. 
It is highly non-trivial to impose macroscopically inspired boundary
conditions on such microscopic codes, see e.g.\ \cite{LiYip} for a control-based 
strategy.  
This can be circumvented by introducing buffer regions at the boundary
of each small box, which shield the short-time dynamics within the computational
domain of interest from boundary effects.  One then uses
 the microscopic code with its \emph{built-in} boundary conditions.     

Here, we show we can study the gap-tooth scheme (with buffers) 
through its numerically obtained damping factors,
by estimating its eigenvalues.  
Integration with nearby coarse initial conditions is used to
estimate matrix-vector products of the linearization of the coarse time-$\Delta t$ map
with known perturbation vectors; these are integrated in matrix-free iterative
methods such as Arnoldi eigensolvers.
For a standard diffusion problem,
we show that the eigenvalues of the gap-tooth scheme are approximately 
the same as those of the finite difference scheme.  When we impose 
Dirichlet boundary conditions at the boundary of the buffers, we show
that the scheme converges to the standard gap-tooth scheme for increasing 
buffer size. 

\section{Physical Problem/ Governing Equations}
\label{sec:2}

We consider a general reaction-convection-diffusion equation with
a dependence on a small parameter $\epsilon$,
\begin{equation}\label{eq:general_homogenization_problem}
\frac{\partial}{\partial
t}u(x,t)=f\left(u(x,t),\frac{\partial}{\partial
x}u(x,t),\frac{\partial^2}{\partial
x^2}u(x,t),x,\frac{x}{\epsilon} \right),
\end{equation}
with initial condition $u(x,0)=u_0(x)$ and Dirichlet boundary
conditions $u(0,t)=v_o$ and $u(1,t)=v_1$.  We further assume that
$f$ is 1-periodic in $y=\frac{x}{\epsilon}$.

Since we are only interested in the macroscopic (averaged)
behavior, let us define an averaging operator for $u(x,t)$ as
follows
\begin{equation}
U(x,t):=\mathcal{S}_h(u)(x,t)=\int_{x-\frac{h}{2}}^{x+\frac{h}{2}}u(\xi,t)\d\xi.
\end{equation}
This operator replaces the unknown function with its local average
in a small box of size $h>>\epsilon$ around each point.  If $h$ is
sufficiently small, this amounts to the removal of the microscopic
oscillations of the solution, retaining its macroscopically
varying components.

The averaged solution $U(x,t)$ satisfies an (unknown) macroscopic
partial differential equation,
\begin{equation}\label{eq:general_macroscopic_equation}
\frac{\partial}{\partial
t}U(x,t)=F\left(U(x,t),\frac{\partial}{\partial
x}U(x,t),\frac{\partial^2}{\partial x^2}U(x,t),x;h \right),
\end{equation}
which does not depend on the small scale, but instead has a dependence on
the box width $h$.

The goal of the gap-tooth scheme is to approximate the solution $U(x,t)$, while
only making use of the detailed 
model (\ref{eq:general_homogenization_problem}).
For illustration purposes, consider as a microscopic model the constant coefficient 
diffusion equation,
\begin{equation}\label{eq:diffusion}
\frac{\partial}{\partial t}u(x,t)=a^*\frac{\partial ^2}{\partial x^2}u(x,t),
\end{equation}
In this case both $U(x,t)$ and $u(x,t)$ satisfy (\ref{eq:diffusion}).  
The microscopic and macroscopic models are the same, 
which allows us to focus completely on the method and its implementation.

\section{Multiscale/Multiresolution Method}
\label{sec:3}

\subsection{The gap-tooth scheme\label{sec:3.1}}
 Suppose we want to obtain the solution of the \emph{unknown}
 equation
(\ref{eq:general_macroscopic_equation}) on the interval $[0,1]$,
using an equidistant, macroscopic mesh $\Pi(\Delta
x):=\{0=x_0<x_1=x_0+\Delta x<\ldots<x_N=1\}$. To this end,
consider a small interval (\emph{tooth}, box) of length $h<<\Delta
x$ centered around each mesh point, and let us perform a time integration
using the microscopic model
 (\ref{eq:general_homogenization_problem}) in each box. We provide
each box with boundary conditions and initial condition as follows.

{\bf Boundary conditions.} Since the microscopic model
(\ref{eq:general_homogenization_problem}) is diffusive, it makes sense to 
impose a fixed gradient at the boundary of each small box
for a time $\Delta t$ for the macroscopic function $U(x,t)$. The value of this 
gradient is determined by
an approximation of the concentration profile by a polynomial, based
on the (given) box averages  $U_i^n$, $i=1,\ldots, N$.
\begin{displaymath}
u(x,t_n)\approx p_i^k(x;t_n),\qquad x \in
[x_i-\frac{h}{2},x_i+\frac{h}{2}],
\end{displaymath}
where $p_i^k(x;t_n)$ denotes a polynomial of (even) degree $k$.
We require that
the approximating polynomial has the same box averages in box $i$
and in $\frac{k}{2}$ boxes to the left and to the right.  This gives us
\begin{equation}\label{eq:polynomial_condition}
\frac{1}{h}\int_{x_{i+j}-\frac{h}{2}}^{x_{i+j}+\frac{h}{2}}
p_i^k(\xi;t_n)\d\xi=U_{i+j}^n, \qquad
j=-\frac{k}{2},\ldots,\frac{k}{2}.
\end{equation}
One can easily check
that
\begin{equation}\label{eq:lagrange}
\mathcal{S}_{h}(p_i^k)(x,t_n)=\sum_{j=-\frac{k}{2}}^{\frac{k}{2}}U_{i+j}^n
L_{i,j}^k(x), \qquad
L_{i,j}^k(x)=\prod_{\stackrel{l=-\frac{k}{2}}{ l\ne
j}}^{\frac{k}{2}}\frac{(x-x_{i+l})}{(x_{i+j}-x_{i+l})}
\end{equation}
where $L_{i,j}^k(x)$ denotes a
Lagrange polynomial of degree $k$.
The derivative of $p_i^k(x,t_n)$ at $x_i\pm\frac{h}{2}$ is subsequently
used as a Neumann boundary condition,
\begin{equation}\label{eq:bc}
\left.\frac{\d}{\d x}p_i^k(x;t_n)\right|_{x_i\pm\frac{h}{2}}
\end{equation}
In \cite{SamKevrRoo03}, it is shown how to enforce the
macroscopic gradient to be constant when the microscopic
model exhibits fast oscillations. 

{\bf Initial condition.}
For the time integration, we must impose an initial condition 
$\tilde{u}^i(x,t_n)$ 
in each box $[x_i-\frac{h}{2},x_i+\frac{h}{2}]$, at time $t_n$.  
We require
$\tilde{u}^i(x,t_n)$ to satisfy the boundary condition and 
the given box average.
We choose a quadratic polynomial,
centered around the coarse mesh point,
\begin{equation}\label{eq:lifting_quad}
\tilde{u}^i(x,t_n)\equiv a(x-x_i)^2+b(x-x_i)+c.
\end{equation}
Using the constraints (\ref{eq:bc}) 
and requiring $\frac{1}{h}\int_{x_{i}-\frac{h}{2}}^{x_i+\frac{h}{2}}\tilde{u}^i(\xi,t_n)\d\xi
=U_i^n$,
 we obtain
%\begin{eqnarray}
%a&=&\frac{s_i^+ -s_i^-}{2h},\notag\\
%b&=&\frac{s_i^+ +s_i^-}{2},\label{eq:lifting}\\
%c&=&{U}_i^n-\frac{h^2}{24}(s_i^+ - s_i^-)\notag.
%\end{eqnarray}
\begin{equation}
a=\frac{s_i^+ -s_i^-}{2h},\qquad b=\frac{s_i^+ +s_i^-}{2}, \qquad
c={U}_i^n-\frac{h}{24}(s_i^+ - s_i^-)\label{eq:lifting}.
\end{equation}

{\bf The algorithm.}  The complete \emph{gap-tooth} algorithm
to proceed from $t_n$ to $t_{n+1}=t_n+\Delta t$ is given below:
\begin{enumerate}
\item At time $t_n$, construct
the initial condition $\tilde{u}^i(x,t_n)$, $i=0,\ldots,N$, 
using the box averages $U^n_j$ ($j=0,\ldots, N$) as defined in 
(\ref{eq:lifting}).
\item Compute $\tilde{u}^i(x,t)$ by solving the equation
(\ref{eq:general_homogenization_problem}) until time
$t_{n+1}=t+\Delta t$ with Neumann boundary conditions (\ref{eq:bc}).
\item Compute the box average $U_i^{n+1}$ at time $t_{n+1}$.
\end{enumerate}

It is clear that this amounts to a ``coarse to coarse'' time-$\Delta t$ map.
We write this map as follows,
\begin{equation}
U^{n+1}=S_k(U^n;t_n+\Delta t),
\end{equation}
where $S$ represents the numerical time-stepping scheme for the
macroscopic (coarse) variables and $k$ denotes the degree of
interpolation.

 We emphasize that the
scheme is also applicable if the microscopic model is not a
partial differential equation. In this case, we replace step 2
with a coarse time-stepper, based on the \emph{lift-run-restrict}
procedure that was outlined in \cite{GearKevrTheo02}.
Numerical experiments using this algorithm are presented in
\cite{GearLiKevr03,GearKevr02}. Figure \ref{fig:schematic} gives a
schematic representation of the algorithm.

\subsection{The gap-tooth scheme with buffers}\label{sec:3.2}

\begin{figure}
\centering
\subfigure[The gap-tooth scheme\label{fig:schematic}]{
\includegraphics[scale=0.25]{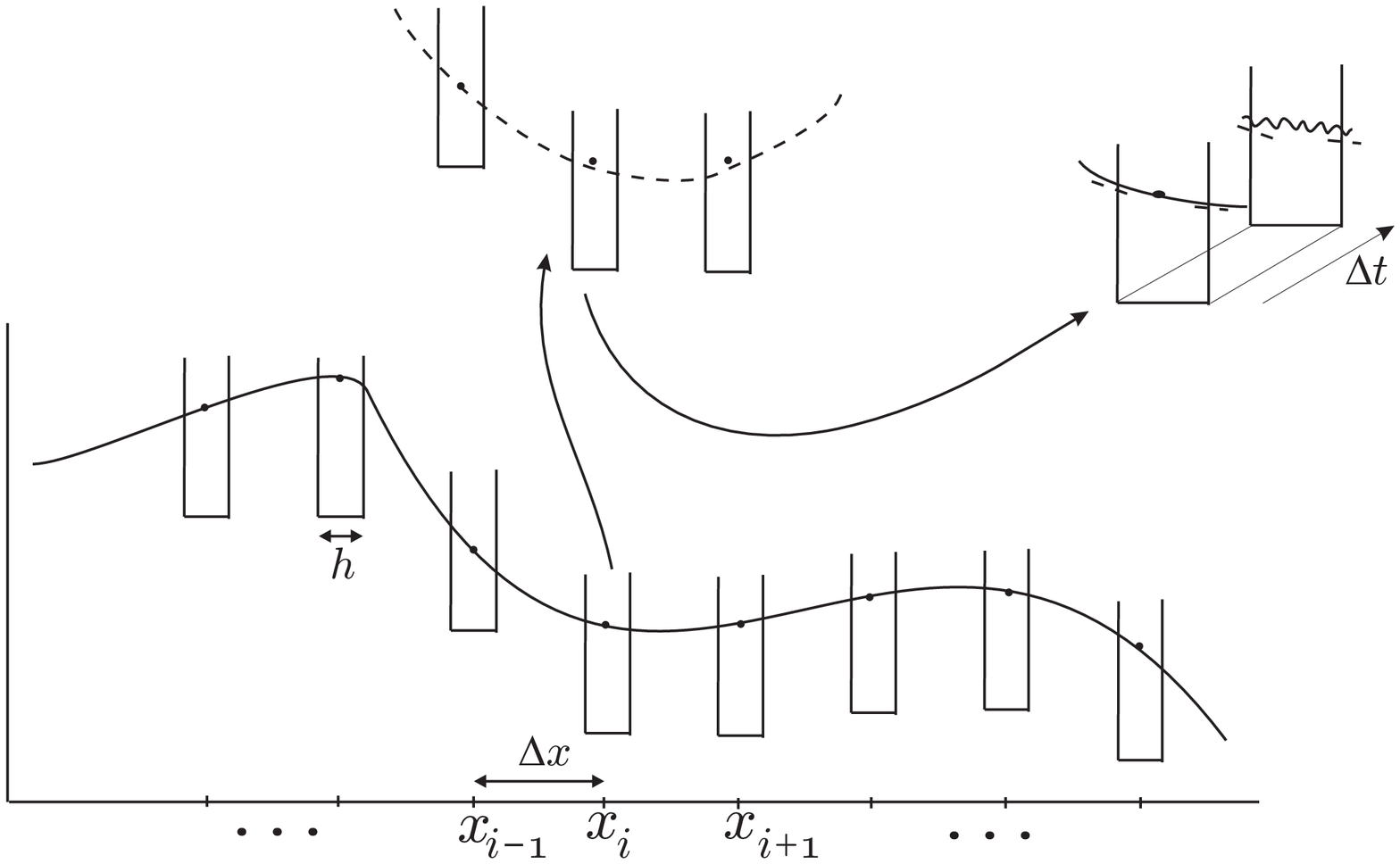}}
\subfigure[Introduction of buffers\label{fig:schematic_buffer}]{
\includegraphics[scale=0.25]{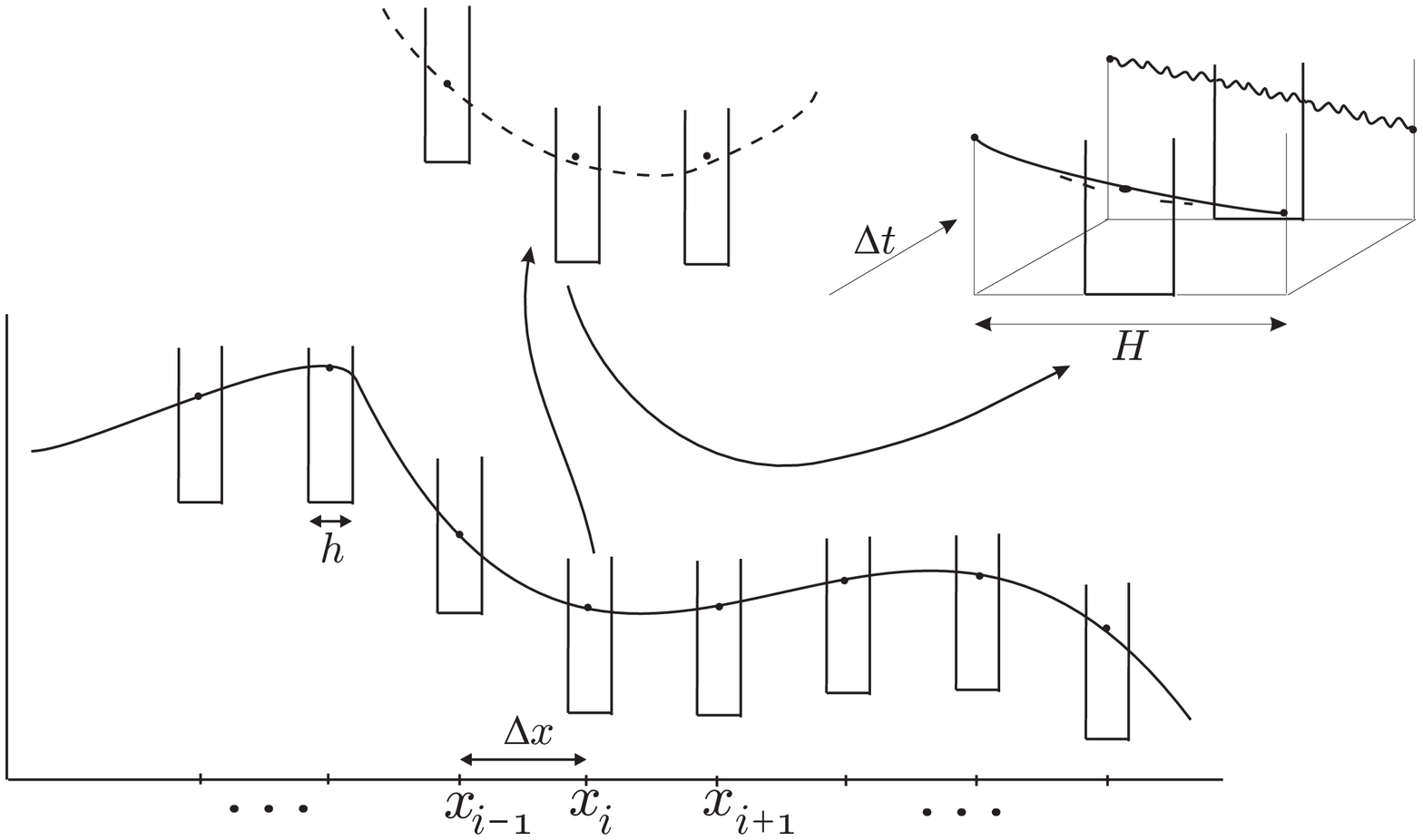}}
\caption{\label{fig:schematic_both}A schematic representation of
the two variants of the gap-tooth scheme (with and without buffer boxes).
%gap-tooth initialization.  We choose a number of boxes of size $h$
%around each macroscopic mesh point $x_i$ and interpolate the
%initial averages (dots) in a number of boxes around $x_i$ (dashed
%profile). The derivatives at the boundary and the average are used
%to create an initial profile in box $i$ (dotted line).
}
\end{figure}

We already mentioned that, in many cases, it is not possible or convenient
to constrain the macroscopic gradient.  
However,
the only crucial issue is that the detailed system in each box should evolve 
\emph{as if it were embedded in a larger domain}.
This can be effectively accomplished by introducing a larger box of size $H>>h$ 
around each
macroscopic mesh point, but still only use (for macro-purposes) the evolution over
the smaller, ``inner" box.
This is illustrated 
in figure \ref{fig:schematic_buffer}.
Lifting and evolution (using \emph{arbitrary} outer boundary
conditions) are performed in the larger box; yet the restriction is done by
taking the average of the solution over the inner, small box.
The goal of the additional computational domains, the \emph{buffers}, is to
buffer the solution inside the small box from outer boundary effects.
This can be accomplished over \emph{short enough} times, provided the buffers are \emph{large enough};
analyzing the method is tantamount to making these statements quantitative.

The idea of using a buffer region was also used in the multi-scale finite element method 
(oversampling) of Hou \cite{HouWu97}
to eliminate the boundary layer effect; also Hadjiconstantinou makes use of overlap 
regions to couple a particle simulator with a continuum code \cite{Hadji99}.  
If the microscopic code allows a choice of
different types of ``outer" microscopic boundary conditions, 
selecting the size of the buffer may
also depend on this choice.

\section{Results}
\label{sec:4}

We first show analytically and numerically that the standard
gap-tooth scheme converges for equation (\ref{eq:diffusion}).
We then analyze convergence of the scheme with buffers and 
Dirichlet boundary conditions through its damping factors.

\subsection{Convergence of the gap-tooth scheme}

\begin{theorem}
The gap-tooth scheme, applied to equation (\ref{eq:diffusion})
with exact, analytical integration
within the boxes, and boundary conditions defined through
interpolating polynomials of (even) order $k$, is equivalent 
to a finite difference discretization of equation (\ref{eq:diffusion}) 
of order
$k$ central differences in space and an explicit Euler time step.
\end{theorem}
\proof
When using exact (analytic) integration in each box, we can find
an explicit formula for the gap-tooth time-stepper.
The initial profile is given by (\ref{eq:lifting_quad}),
$
\tilde{u}_i(x,t_n)=a(x-x_i)^2+b(x-x_i)+c
$.
Due to the Neumann boundary conditions, time integration can be done
analytically, using
\begin{displaymath}
\tilde{u}(x,t_n+\Delta t)=a(x-x_i)^2+b(x-x_i)+c+2a\cdot a^*\Delta t
\end{displaymath}
Averaging this profile over the box gives the
following time-stepper for the box averages,
\begin{displaymath}
U_i^n=U_i^n+a^*\frac{s^+_i-s^-_i}{h}\Delta
t\label{eq:proof_time_stepper}.
\end{displaymath}
We know that
$s^{\pm}_i=\left.\frac{\d}{\d
x}p_i^k(x,t_n)\right|_{x_i\pm\frac{h}{2}},
$
where $p_i^k(x,t_n)$  is determined by (\ref{eq:polynomial_condition}).
%Since $\mathcal{S}_h(p_i^k)(x,t_n)$ is a Lagrange interpolating polynomial of
%order $k$ for the box averages (see equation (\ref{eq:lagrange})),
%we obtain a $k$-th order approximation of the second derivative
%of $U(x,t_n)$ at $x=x_i$ by taking the second derivative of $p_i^k(x,t_n)$. 
%This
%is true for $k$ even, due to symmetry reasons.  
One can easily verify that
\begin{displaymath}
\frac{\d^2}{\d x^2}\mathcal{S}_h(p_i^k)(x,t_n)=\frac{s^+_i-s^-_i}{h},
\end{displaymath} is a $k$-th order approximation of $\frac{\partial^2 u}{\partial x^2}$, which concludes the proof.\qed
\endproof

\begin{figure}
\centering
\subfigure[Numerical solution]
{\includegraphics[scale=0.3]{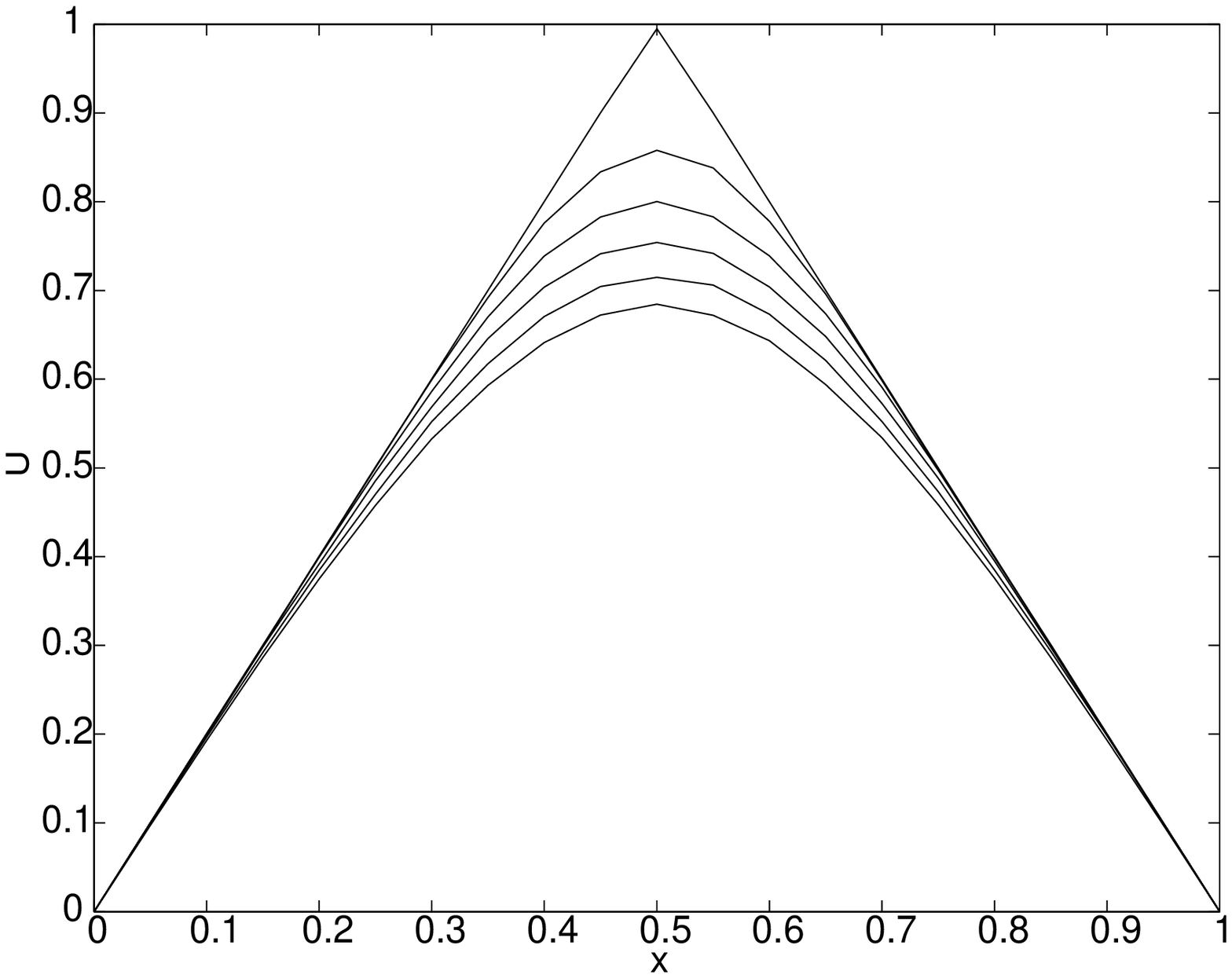}}
\subfigure[Difference with FD]
{\includegraphics[scale=0.3]{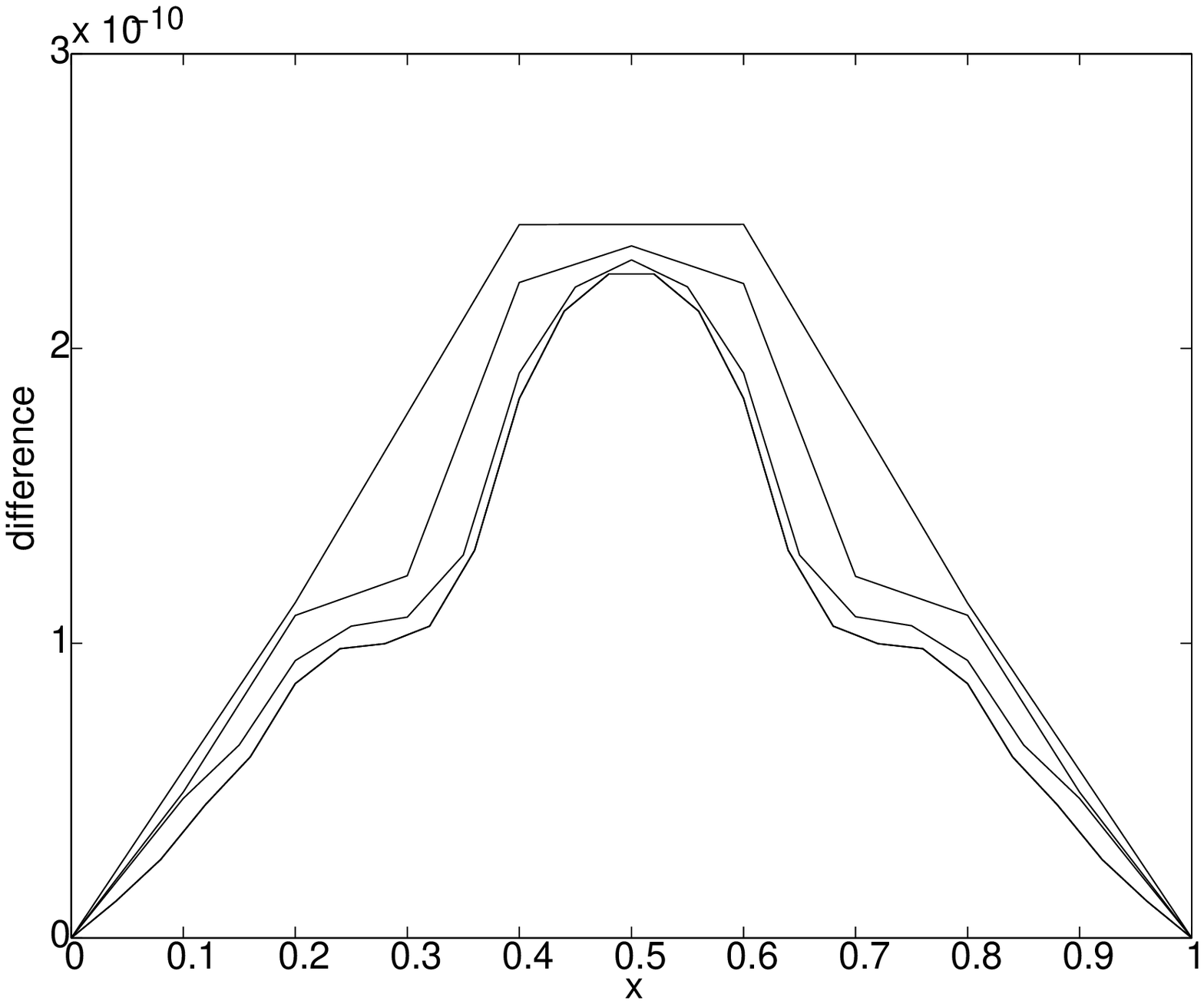}}
\caption{\label{fig:convergence}The gap-tooth scheme of fourth order for eq.\ 
(\ref{eq:diffusion}) at $t=0,4\cdot 10^{-3},\ldots,2\cdot 10^{-2}$. }
\end{figure}

As an example, we apply the gap-tooth scheme to the diffusion equation
(\ref{eq:diffusion}) with $a^*=1$.
We choose an initial condition $U(x,t)=1-|2x-1|$, with Dirichlet boundary conditions,
and show the result of a fourth-order gap-tooth simulation with $\Delta x=0.05$,
$\Delta t=5\cdot 10^{-3}$ and
$h=0.01$.  Inside each box, we used a second order finite difference scheme
 with microscopic
spatial mesh size $\delta x=1\cdot 10^{-3}$
and $\delta t=5\cdot 10^{-7}$. The results are shown in figure \ref{fig:convergence}.

\subsection{Damping factors}

Convergence results are typically established by proving
 consistency and stability.
If one can prove that the error in each time step can be
 made arbitrarily small by
refining the spatial and temporal mesh size, and that an
 error made at time $t_n$ does
not get amplified in future time-steps, one has proved 
convergence.  This requires the solution operator to be 
stable as well.

In the abscense of explicit formulas, one can examine
 the damping factors of the
time-stepper.  If, for decreasing mesh sizes, all 
(finitely many) eigenvalues
and eigenfunctions of the
time-stepper converge to the
dominant eigenvalues and eigenfunctions of the 
time evolution operator, one expects the solution
of the scheme to converge to the true solution of the evolution problem.

%{\bf we need something about stability here ! the eigenvalues must be stable ?}.

Consider equation (\ref{eq:diffusion}) with Dirichlet boundary
conditions $u(0,t)=0$ and $u(1,t)$, and denote its solution at
time $t$ by the time evolution operator
\begin{equation}\label{eq:evolution_operator}
u(x,t)=s(u_0(x);t),
\end{equation}
We know that
\begin{displaymath}
s(\sin(m\pi x);t)=\e^{-(m\pi)^2 t} \sin(m\pi x), \qquad m\in
\mathbb{N}.
\end{displaymath}
Therefore, if we consider the time evolution operator over a fixed
time $\bar{t}$, $s(\cdot,\bar{t})$, then this operator has
eigenfunctions $\sin(m\pi x)$, with resp.\ eigenvalues
\begin{equation}\label{eq:dispersion_relation}
\lambda_m=\e^{-{(m\pi)}^2\bar{t}}.
\end{equation}  
A good (finite
difference) scheme approximates well all
eigenvalues whose eigenfunctions can be
represented on the given mesh. We note that it is possible
to decouple the time horizon $\bar{t}$ from the gap-tooth 
(or finite-difference) time-step $\Delta t$, in order to study the effect of different
discretizations on the same reporting horizon.

Since the operator defined in (\ref{eq:evolution_operator})
 is linear, the 
numerical time integration is equivalent to a 
matrix-vector product.  Therefore,
we can compute the eigenvalues using matrix-free
linear algebra techniques, even for the gap-tooth scheme, 
for which it might not even be 
possible to obtain a closed
expression for the matrix.
We note that this analysis gives us an indication about the 
quality of the scheme, but it is by no means a proof of 
convergence.

\begin{figure}
\centering
\subfigure{\includegraphics[scale=0.3]{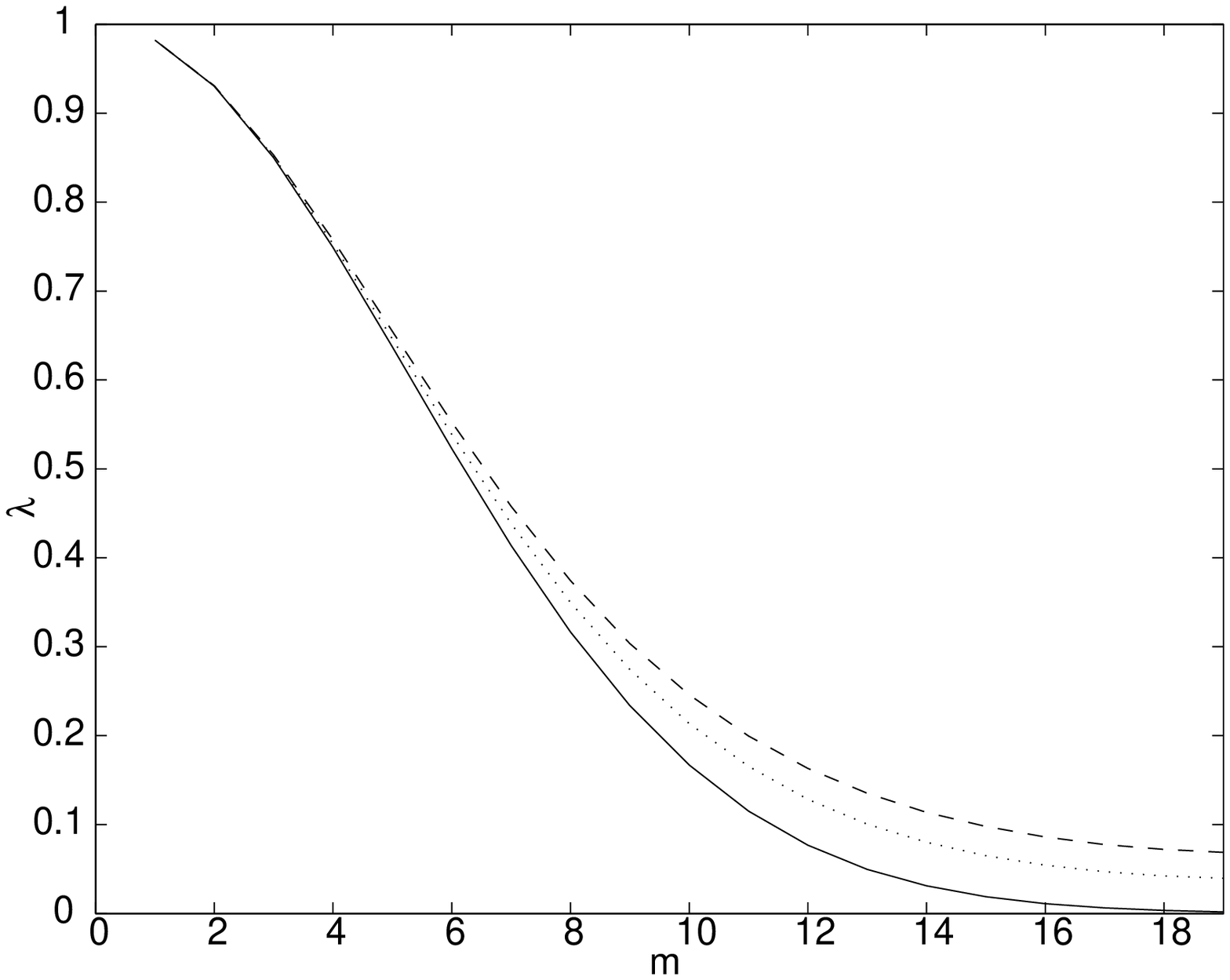}}
\subfigure{\includegraphics[scale=0.3]{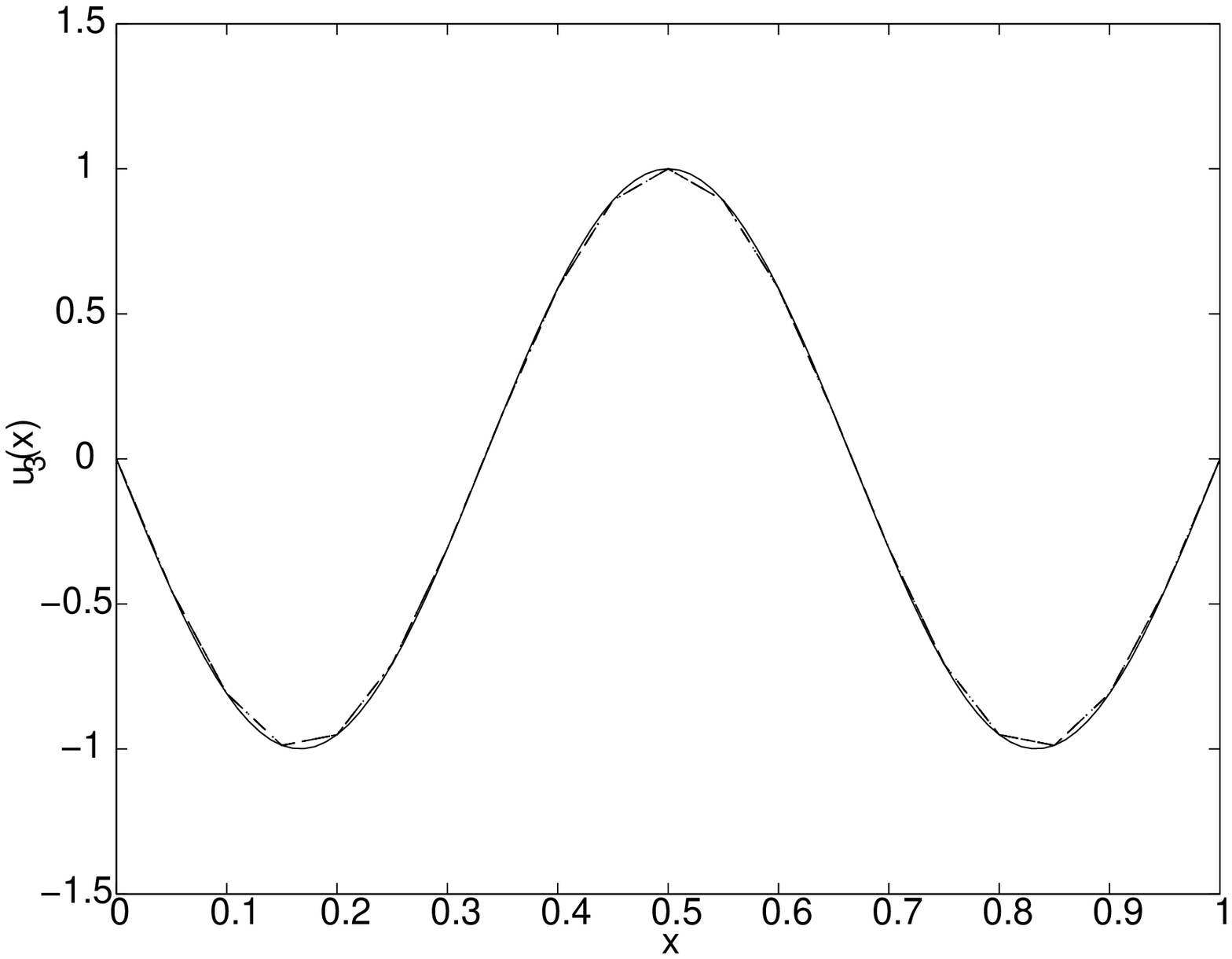}}
\caption{\label{fig:eigval_neu}Comparison between 
the damping factors (left) and the eigenfunction
corresponding to eigenvalue $\lambda_3$ (right) of the 
exact solution (full line), the finite difference
approximation (dashed) and the gap-tooth scheme (dotted). 
The eigenfunction of the gap-tooth scheme is indistinguishable
of the finite difference eigenfunction.}
\end{figure}

We illustrate this with the computation of the eigenvalues of the
gap-tooth scheme with Neumann box boundary conditions.  In this case,
we know from theorem 1 that these eigenvalues should correspond to
the eigenvalues of a finite difference scheme on the same mesh. We
compare the eigenvalues of the gap-tooth scheme of order $k=2$ 
for equation (\ref{eq:diffusion}) with
diffusion coefficient $a^*=0.45825686$. As method parameters, 
we choose $\Delta x=0.05$, 
$h=5\cdot 10^{-3}$, 
$\Delta t=2.5\cdot 10^{-4}$ for a time horizon 
$\bar{t}=4\cdot 10^{-3}$, which corresponds to 16 gap-tooth steps.  
Inside each box,
we use a finite difference scheme of order 
$2$ with $\delta x=1\cdot 10^{-4}$ and an implicit
Euler time-step of $5\cdot 10^{-5}$. 
We compare
these eigenvalues to those the finite difference 
scheme with $\Delta x=0.05$ and 
$\Delta t=2.5\cdot 10^{-4}$, and with the dominant 
eigenvalues of the ``exact'' solution
(a finite difference approximation with 
$\Delta x=1\cdot 10^{-3}$ and $\Delta t=1\cdot 10^{-7}$).
The result is shown in figure \ref{fig:eigval_neu}.  
The difference between the finite difference
approximation and the gap-tooth scheme in the higher modes, 
which should be zero according to 
theorem 1, is due to the numerical solution inside each box and the use
of numerical quadrature for the average.

\begin{figure}
\centering
\subfigure{\includegraphics[scale=0.3]{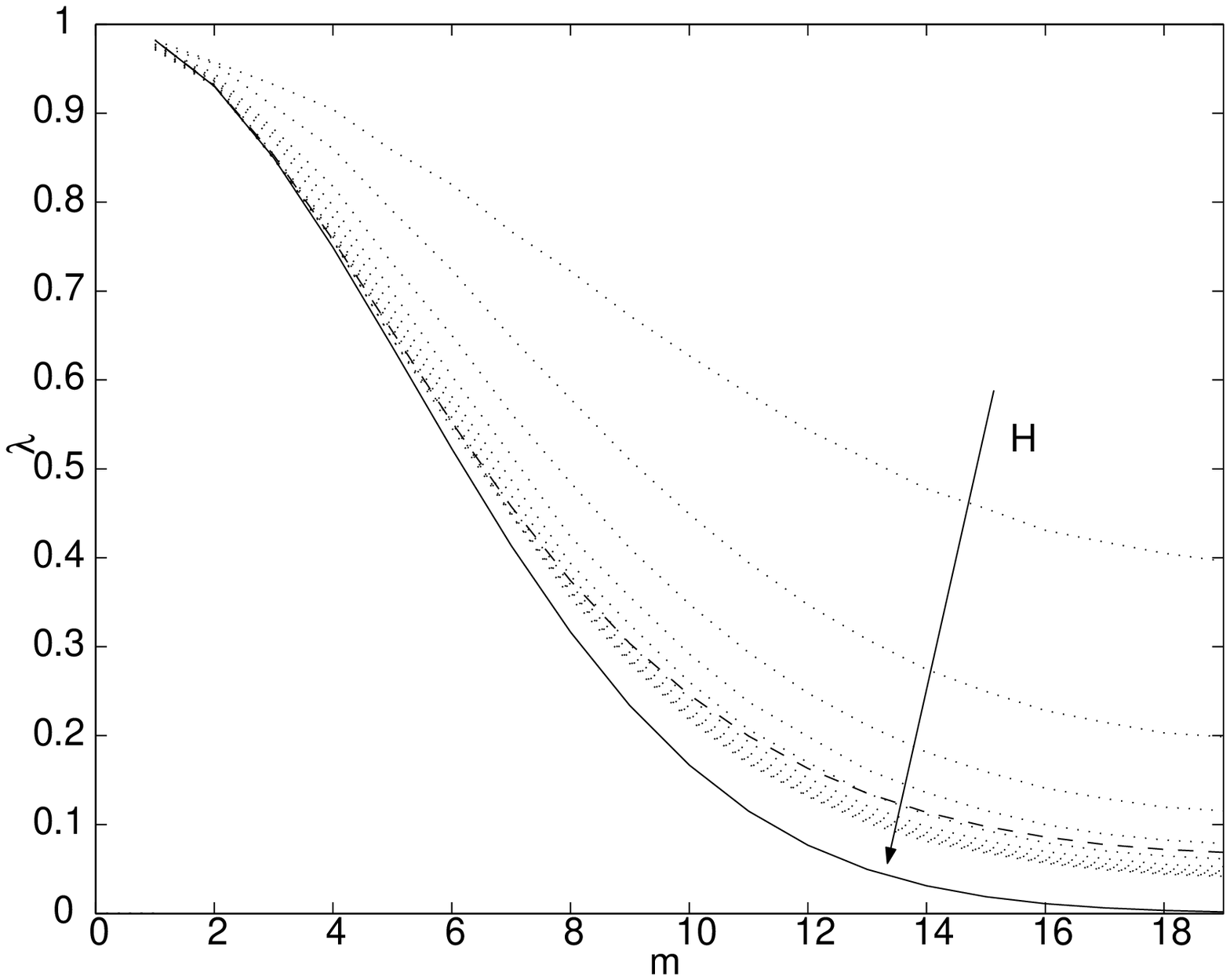}}
\subfigure{\includegraphics[scale=0.3]{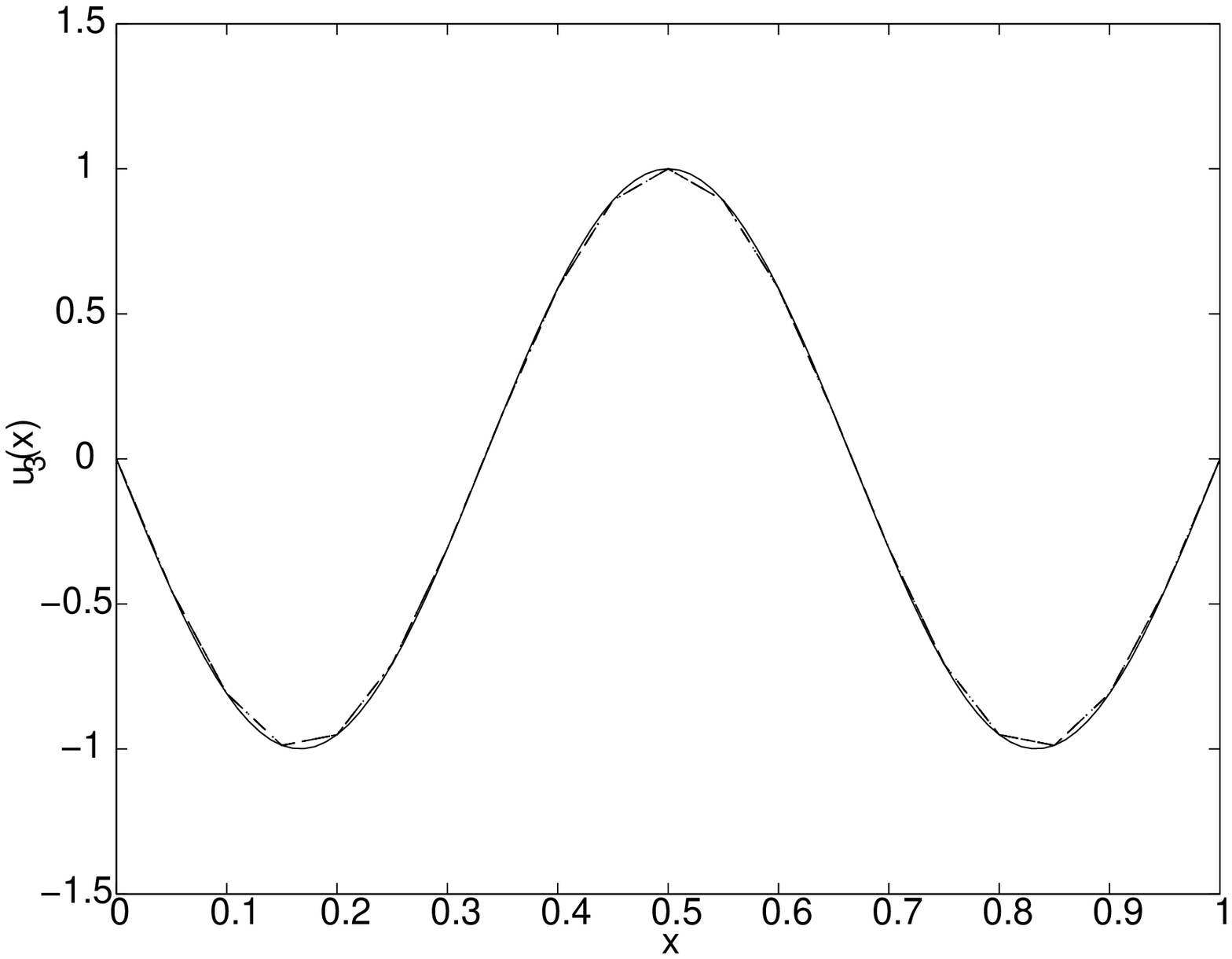}}
\caption{\label{fig:eigval_dir}Comparison between the damping factors (left) and
the eigenfunction corresponding to the eigenvalue $\lambda_3$ (right) of
the exact solution (full line), the finite difference scheme (dashed) and 
the gap-tooth scheme with buffers (dash-dotted lines) for increasing buffer sizes
$H=2\cdot 10^{-2},3\cdot 10^{-2}\ldots,1\cdot 10^{-1}$.}
\end{figure}

We now examine the effect of introducing a buffer region, 
as described in section \ref{sec:3.2}.
We consider again equation  (\ref{eq:diffusion}) 
with $a^*=0.45825686$, and we take the gap-tooth 
scheme with parameters  $\Delta x=0.05$, 
$h=5\cdot 10^{-3}$, 
$\Delta t=2.5\cdot 10^{-4}$ for a time horizon 
$\bar{t}=4\cdot 10^{-3}$, and an internal time-stepper as above.  
We introduce a buffer region of size $H$, and we impose 
Dirichlet boundary conditions at the outer boundary
of the buffer region.  Lifting is done in identically the same way as 
for the gap-tooth scheme without buffers; we only use (\ref{eq:lifting}) 
as the initial condition in the larger box $[x_i-\frac{H}{2},x_i+\frac{H}{2}]$. 
We compare the eigenvalues again with the 
equivalent finite difference scheme and 
the exact solution, for increasing sizes of the buffer box $H$.  
Figure \ref{fig:eigval_dir} shows that, as $H$ increases, the eigenvalues
of the scheme
converge to those of the original gap-tooth scheme.   We see that,
in this case, we would need a buffer of size $H=4\cdot 10^{-2}$, i.e.\ 80\% 
of the original domain, for a good approximation of the damping factors.   
Note that it is conceptually
possible, though inefficient, to let the buffer boxes overlap.

\section{Summary/Conclusions}
\label{sec:5}
We described the gap-tooth scheme for the numerical simulation of 
multi-scale problems.  This scheme simulates the macroscopic behaviour 
over a macroscopic domain when only a microscopic model is explicitly 
available.  
In the case of diffusion, we showed equivalence of our
scheme to standard finite differences of arbitrary (even) order, 
both theoretically
and numerically.  

 We showed that it is possible, even without
analytic formulas, to study the properties
of the gap-tooth scheme and generalizations through the damping
factors of the
resulting coarse time-$\Delta t$ map.  
We illustrated this for the original gap-tooth scheme and 
for an implementation
using Dirichlet boundary conditions in a buffer box.   We showed that, 
as long
as the buffer region is ``large enough'' to shield the internal region 
from the boundary effects over a time $\Delta t$,
we get a convergent scheme.  Therefore, we are able to use microscopic codes 
in the gap-tooth scheme
\emph{without modification}.

In a forthcoming paper, we will explore, using these damping factors 
for many different types of boundary conditions, the 
relation between
the quality of the boundary conditions, the size of the buffer 
region and the time-step
before reinitialization.  We will investigate the trade-off between
the effort required to impose a particular type of boundary conditions (and the eventual
macroscopically inspired control-based strategy) and the efficiency gain
due to smaller buffer sizes and/or longer possible time-steps before
reinitialization.  Here, we showed that this investigation is made possible 
by studying the damping factors of the resulting coarse time-$\Delta t$ map.

%  This should also provide information that allows us
%to check whether the chosen buffer size was large enough.

%{\bf Giovanni, when you first talk about buffers, you should mention at least
%ONE paper of Tom Hou and ONE paper of Nicolas Hadjiconstantinou that use buffers
%for multiscale problems --
%I am attaching two such references for the latter, choose one}

{\bf Acknowledgements}  GS is a Research Assistant of the Fund of 
Scientific Research - Flanders.  This work has been partially supported
by an IUAP grant and by the Fund of Scientific Research 
through Research Project G.0130.03 (GS, DR), 
and by the AFOSR and the NSF (IGK).  The authors
thank Olof Runborg for discussions that improved this text and the organizers
of the Summer School in Multi-scale Modeling and Simulation in Lugano.

%
%
% BibTeX users please use
% \bibliographystyle{}
% \bibliography{}

\begin{thebibliography}{[KLR73]}
% Use the following syntax and markup for your references
%
% Monographs
\bibitem[EE03]{EEng03}
W.~E and B.~Engquist.
The heterogeneous multi-scale methods. Comm. Math. Sci., 1(1):87--132, 2003.
\bibitem[GK02]{GearKevr02}
C.W. Gear and I.G. Kevrekidis. Boundary processing for {M}onte
{C}arlo simulations in the gap-tooth scheme. physics/0211043 at
arXiv.org, 2002.
\bibitem[GKT02]{GearKevrTheo02}
C.W. Gear, I.G. Kevrekidis, and C.~Theodoropoulos.
``{C}oarse'' integration/bifurcation analysis via microscopic
  simulators: micro-{G}alerkin methods. Comp. Chem. Eng., 26(7-8):941--963,2002.
\bibitem[GLK03]{GearLiKevr03}
C.W. Gear, J.~Li, and I.G. Kevrekidis. The gap-tooth method in
particle simulations. Physics Letters A, 316:190--195, 2003.
\bibitem[Had99]{Hadji99} N. G. Hadjiconstantinou. Hybrid
atomistic-continuum formulations and the moving contact-line
problem. J. Comp. Phys., 154:245--265, 1999.
\bibitem[HW97]{HouWu97}
T.Y. Hou and X.H. Wu.
A multiscale finite element method for elliptic problems in composite
  materials and porous media.
J.\ Comp.\ Phys.\ , 134:169--189, 1997.
\bibitem[KGK02]{KevrGearHymKevrRunTheo02}
I.G. Kevrekidis, C.W. Gear, J.M. Hyman, P.G. Kevrekidis, O.~Runborg, and
  C.~Theodoropoulos. Equation-free multiscale computation:
enabling microscopic simulators
  to perform system-level tasks. Comm. Math. Sci. Submitted, physics/0209043 at arxiv.org, 2002.
\bibitem[LLY98]{LiYip}
J. Li, D. Liao, S. Yip. Imposing field boundary conditions
in MD simulation of fluids: optimal particle controller and
buffer zone feedback, Mat. Res. Soc. Symp. Proc, 538(473-478), 1998.
\bibitem[SKR03]{SamKevrRoo03}
G. Samaey, I.G. Kevrekidis, and D. Roose. The gap-tooth scheme for
homogenization problems. SIAM MMS, 2003. Submitted.
\bibitem[SGK03]{SetGearOthKevr03}
S.~Setayeshar, C.W. Gear, H.G. Othmer, and I.G. Kevrekidis.
Application of coarse integration to bacterial chemotaxis. SIAM
MMS. Submitted, physics/0308040 at arxiv.org, 2003.
\end{thebibliography}
%
% Non-BibTeX users please follow the syntax
% the syntax of "referenc.tex" for your own citations

%%%%%%%%%%%%%%%%%%%%%%%%%%%%%%%%%%%%%%%%%%%%%%%%%%%%%%%%%%%%%%%%%%%%%%

%%%%%%%%%%%%%%%%%%%%%%%%%%%%%%%%%%%%%%%%%%%%%%%%%%%%%%%%%%%%%%%%%%%%%%

\printindex
\end{document}